\documentclass[reprint,aps,prb,superscriptaddress,secnumarabic,amssymb, nobibnotes]{revtex4-1}
\usepackage{graphicx}
\usepackage{amsmath}

\begin{document}

\title{Investigating optically-excited THz standing spin waves using noncollinear magnetic bilayers}

\author{M.L.M. Lalieu}
\email[Corresponding author: ]{m.l.m.lalieu@tue.nl}
\affiliation{Department of Applied Physics, Institute for Photonic Integration, Eindhoven University of Technology, P.O. Box 513, 5600 MB Eindhoven, The Netherlands}
\author{R. Lavrijsen}
\affiliation{Department of Applied Physics, Institute for Photonic Integration, Eindhoven University of Technology, P.O. Box 513, 5600 MB Eindhoven, The Netherlands}
\author{R.A. Duine}
\affiliation{Institute for Theoretical Physics, Utrecht University, Princetonplein 5, 3584 CC Utrecht, The Netherlands}
\affiliation{Department of Applied Physics, Institute for Photonic Integration, Eindhoven University of Technology, P.O. Box 513, 5600 MB Eindhoven, The Netherlands}
\author{B. Koopmans}
\affiliation{Department of Applied Physics, Institute for Photonic Integration, Eindhoven University of Technology, P.O. Box 513, 5600 MB Eindhoven, The Netherlands}
\date{\today}
\begin{abstract}
We investigate optically excited THz standing spin waves in noncollinear magnetic bilayers. Using femtosecond laser-pulse excitation, a spin current is generated in the first ferromagnetic (FM) layer, and flows through a conductive spacer layer to be injected into the second (transverse) FM layer, where it exerts a spin-transfer torque on the magnetization and excites higher-order standing spin waves. We show that the noncollinear magnetic bilayer is a convenient tool that allows easy excitation of THz spin waves, and can be used to investigate the dispersion and thereby the spin wave stiffness parameter in the thin-film regime. This is experimentally demonstrated using wedge-shaped Co and CoB (absorption) layers. Furthermore, the damping of these THz spin waves is investigated, showing a strong increase of the damping with decreasing absorption layer thickness, much stronger than expected from interface spin pumping effects. Additionally, a previously unseen sudden decrease in the damping for the thinnest films is observed. A model for the additional damping contribution incorporating both these observations is proposed.
\end{abstract}
\maketitle

\section{Introduction}
About a decade ago, it was discovered that spin currents are generated upon femtosecond (fs) laser-pulse excitation of a ferromagnetic (FM) thin film. This was first discovered in a collinear magnetic bilayer, in which the laser-induced transfer of angular momentum between the two FM layers was demonstrated by their influence on the demagnetization dynamics in both layers \cite{Malinowski2008}. In the years that followed, several experiments have demonstrated the direct measurement of the optically excited spin current in a FM/NM (non-magnetic metal) bilayer. In these experiments, the spin current is generated by laser-pulse excitation of the FM, and is detected at the outer NM surface \cite{Mentink2012,Choi2014,Choi2014-2,Hofherr2017}. One of the motivations for the research into the laser-pulse-excited spin current is its potential use in the field of spintronics, in which (electrical) spin currents are already heavily used to manipulate magnetic information in future magnetic data storage devices \cite{Kent2015,Parkin2015}. The manipulation of the magnetization can be pushed to the ultrafast time scale by using the optically generated spin currents. This was demonstrated in recent years using noncollinear magnetic bilayers, in which the laser-induced spin current excited in one FM layer was used to exert a spin-transfer torque (STT) on a second, transversely magnetized, FM layer \cite{Schellekens2014,Choi2014,Choi2015,Balaz2018}. Moreover, it has been demonstrated that the optically excited spin current is absorbed very locally near the injection interface \cite{Lalieu2017}, which allowed the excitation of THz standing spin waves \cite{Razdolski2017,Lalieu2017,Ulrichs2018}. This shows that in addition to its general importance in the field of spintronics, the optically excited spin currents could also be of high potential for future THz magnonics.

In this paper, it is experimentally demonstrated that the noncollinear magnetic bilayer is a convenient tool to generate and investigate optically-excited THz spin waves. Using a wedge-shaped absorption layer (Co or CoB), it is shown that the dispersion and thereby the spin wave stiffness parameter is easily accessible for magnetic layer thicknesses down to a few nanometers. Additionally, the structure allows the investigation of the damping of the THz spin waves and its dependence on the film thickness. The measured damping behavior shows a strong increase of the damping as the layer thickness decreases down to $\approx 10$ nm, which is attributed to the inhomogeneous nature of the spin waves. Moreover, a previously unseen reduction of the damping is seen upon further decrease of the layer thickness. A model describing the observed damping behavior is proposed. For the analysis of the THz standing spin waves, the effective magnetization and Gilbert damping parameter (bulk and interface spin pumping contributions) are needed as a function of the thickness of the absorption layer. These properties are determined using the homogeneous (fundamental) precession, of which the analysis will be discussed first.

\section{Sample structure and characterization}
The basic structure of the noncollinear magnetic bilayers used in this work is given by Si:B(substrate)/Ta(4)/Pt(4)/[Co(0.2)/Ni(0.6)]$_{4}$/Co(0.2) /Cu(5)/FM$_{\mathrm{IP}}$/Pt(1) (thickness in nm), in which two different wedge-shaped in-plane (IP) magnetized (top) FM layers are used; a Co wedge ranging from $0$ to $20$ nm, and a Co$_{77}$B$_{23}$ wedge ranging from $0$ to $15$ nm. These samples are referred to as the Co and the CoB sample in the following. The bottom FM layer is an out-of-plane (OOP) magnetized Co/Ni multilayer. The two FM layers are separated by a $5$nm-thick Cu spacer layer which allows for the transfer of spin currents and decouples both FM layers. All samples are fabricated using dc magnetron sputtering at room temperature. The measurements are performed using a standard time-resolved magneto-optic Kerr effect setup in the polar configuration. The probe and pump pulses have a spot size of $\approx 10 \mu$m and a pulse length of $\approx 150$ fs. The pulses are produced by a Ti:sapphire laser with a wavelength of $790$ nm and a repetition rate of $80$ MHz. During the experiments, the pump pulse excites the spin dynamics, and the probe pulse measures the OOP magnetization component of both FM layers. In case of the homogeneous precession measurements, an external magnetic field is applied parallel to the sample surface.

\begin{figure}
	\includegraphics[scale=0.35]{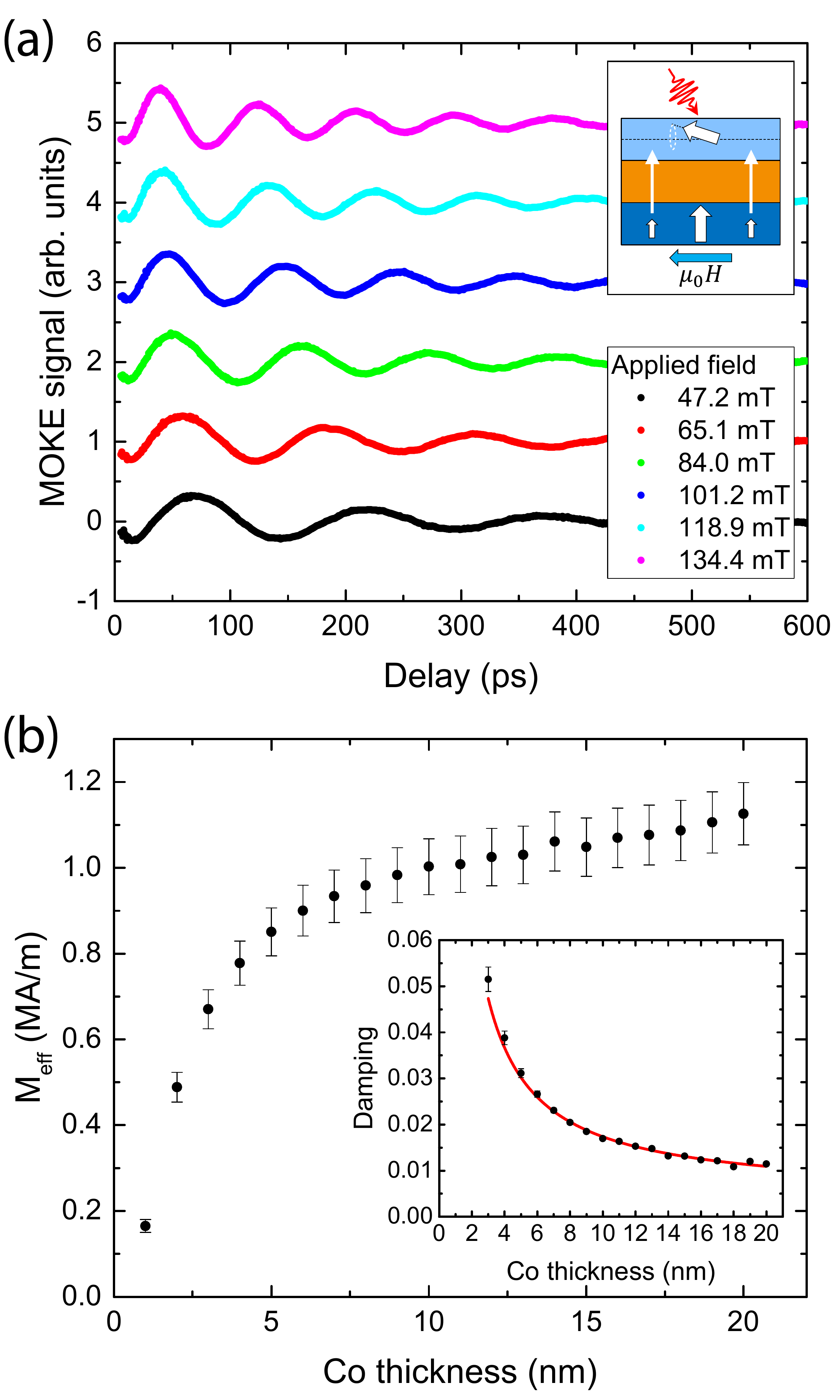}
	\caption{(a) Time-resolved MOKE measurement on the Co sample at a Co thickness of $3$ nm. The figure shows the homogeneous precession for six different magnetic field amplitudes. The background (remagnetization) signal is subtracted, and an offset is added for clarity. The inset shows an illustration of the precession excitation mechanism, based on the ultrafast laser-induced STT. (b) Effective magnetization $M_{\mathrm{eff}}$ as a function of the Co thickness. The inset shows the Gilbert damping parameter as a function of the Co thickness, in which the damping determined with the different magnetic field amplitudes are averaged. The red curve represents a fit to the data using Eq.\ \ref{Eq:Damping}.}
	\label{Fig:KittelPrecessions}
\end{figure}

The effective magnetization $M_{\mathrm{eff}}$ of the IP (absorption) layer at a certain thickness is determined by measuring the frequency $f_{\mathrm{IP}}$ of the homogeneous (fundamental) precession as a function of the applied magnetic field $B_{\mathrm{app}}$. The value of $M_{\mathrm{eff}}$ is obtained by fitting the field dependent frequency using the standard Kittel equation for IP magnetized layers,
\begin{equation}
f_{\mathrm{IP}}=\frac{\gamma}{2 \pi}\sqrt{B_{\mathrm{app}}\left(B_{\mathrm{app}}+\mu_{0}M_{\mathrm{eff}}\right)}.
\label{Eq:FreqKittel}
\end{equation}
In this equation, $\gamma$ corresponds to the gyromagnetic ratio.

The excitation mechanism of the homogeneous precession is the same ultrafast STT mechanism as used for the standing spin wave excitation presented later, and is illustrated in the inset of Fig.\ \ref{Fig:KittelPrecessions}(a). In this mechanism, a fs laser pulse is used to excite a spin current in the OOP (generation) layer. This spin current flows through the Cu spacer layer and is injected into the top FM layer, exerting a STT on the IP magnetization. As a result, the IP magnetization is canted slightly OOP, whereafter it starts a damped precession around the IP applied magnetic field. A more detailed characterization and validation of the excitation mechanism can be found in Refs. \cite{Schellekens2014,Lalieu2017}.

A measurement of the homogeneous precessions in the Co sample at a thickness of $t_{\mathrm{Co}} = 3$ nm, and for six different IP magnetic field amplitudes, is shown in Fig.\ \ref{Fig:KittelPrecessions}(a). The background (remagnetization) signal is subtracted, and an offset is added to the signal for clarity. A clear increase in the precession frequency with the applied field amplitude is observed, as is expected from the Kittel relation. The precessions are fitted using a damped sine, from which the precession frequency $f_{\mathrm{IP}}$ and the characteristic damping time $\tau$ are obtained. Using Eq.\ (\ref{Eq:FreqKittel}), the effective magnetization at each measured Co thickness is determined by fitting the field dependent precession frequency. Figure\ \ref{Fig:KittelPrecessions}(b) shows the effective magnetization as a function of $t_{\mathrm{Co}}$. The observed thickness dependence of $M_{\mathrm{eff}}$ results from an out-of-plane surface anisotropy, which decreases $M_{\mathrm{eff}}$, and of which the contribution falls off as $t_{\mathrm{Co}}^{-1}$. The obtained thickness dependent $M_{\mathrm{eff}}$ is later used in the analysis of the THz standing spin waves. The Kittel fits also allow the determination of the $g$ factor using the fitted value of $\gamma$. For the Co sample, a $g$ factor of $2.30 \pm 0.06$ was found, which is similar as found in literature \cite{Beaujour2006}.

The damped sine fits of the precession data also provide the characteristic damping time $\tau$. Together with the previously determined value of $M_{\mathrm{eff}}$, the Gilbert damping constant $\alpha$ at each thickness and field can be determined using 
\begin{equation}
\alpha=\left[\gamma\tau\left(B_{\mathrm{app}}+\frac{\mu_{0}M_{\mathrm{eff}}}{2}\right)\right]^{-1}.
\label{Eq:DampKittel}
\end{equation}
The damping as a function of the Co thickness is shown in the inset of Fig.\ \ref{Fig:KittelPrecessions}(b), in which the damping determined with the different magnetic field amplitudes are averaged. The damping shows a clear $t_{\mathrm{Co}}^{-1}$ behavior. This thickness dependence is known to be the result of spin pumping into neighboring layers \cite{Kapelrud2013}, in this case at the Cu/Co and Co/Pt interfaces. The interface spin pumping enhances the damping, and since it is an interface effect it falls off as $t_{\mathrm{Co}}^{-1}$. The damping as a function of thickness is fitted using 
\begin{equation}
\alpha = \alpha_{\mathrm{bulk}} + \alpha_{\mathrm{pump}} = \alpha_{\mathrm{bulk}} + \frac{A_{\mathrm{pump}}}{t},
\label{Eq:Damping}
\end{equation} 
in which $\alpha_{\mathrm{bulk}}$ is the (intrinsic) bulk damping, and $\alpha_{\mathrm{pump}}$ is the interface spin pumping contribution to the damping. The interface spin pumping amplitude $A_{\mathrm{pump}}$ includes the contribution of both interfaces. The fitted values are equal to $\alpha_{\mathrm{bulk}} = (4.5 \pm 0.4) \cdot 10^{-3}$ and $A_{\mathrm{pump}} = (1.29 \pm 0.06) \cdot 10^{-10}$ m. The value for $\alpha_{\mathrm{bulk}}$ agrees well with literature values \cite{Tokac2015}. The value of $A_{\mathrm{pump}}$ can be used to calculate the effective spin-mixing conductance of the interfaces \cite{Tserkovnyak2005}, but due to the complex nature of the used multilayers, this is out of the scope of the presented work. Both values are used later when evaluating the damping of the THz standing spin waves. A similar analysis of the homogeneous precessions for the CoB sample is presented in Supplementary Note 1.

\section{Results}
With the effective magnetization and the damping of the homogeneous precession mode characterized, the THz standing spin waves can be investigated using the same noncollinear magnetic bilayers. The higher-order standing spin waves are excited using the same time-resolved polar MOKE measurement as before. Different from the previous measurements is that there is no external magnetic field applied, which is not needed since the standing spin waves are driven by the exchange interaction. Furthermore, to achieve a better sensitivity of the MOKE signal to the THz spin waves, a quarter-wave plate was added to the probe beam \cite{Schellekens2014-2,Lalieu2017}.

\begin{figure}
	\includegraphics[scale=0.35]{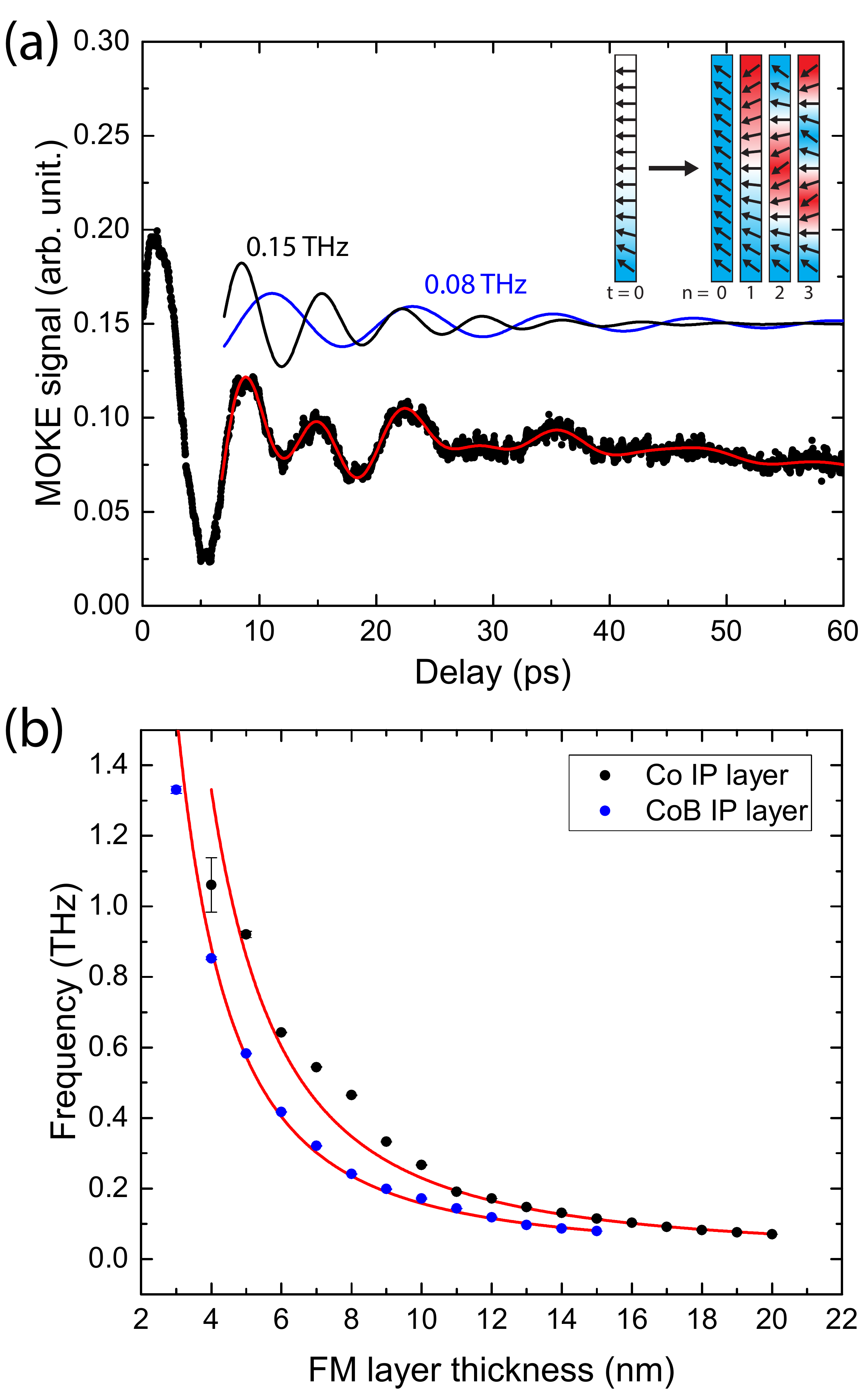}
	\caption{(a) Typical precession measurement of first-order standing spin wave, measured in the Co sample at a Co thickness of $13$ nm. The observed dynamics is a superposition of two damped oscillations, which are illustrated by the blue and black solid lines in the figure. The inset shows an illustration of the excitation mechanism of the standing spin waves. (b) Standing spin wave frequency as a function of the FM layer thickness, for both the Co (black dots) and CoB (blue dots) samples. The red solid lines are fits to the data using Eq.\ \ref{Eq:THzDispersion}.}
	\label{Fig:THzPrecessions}
\end{figure}

An illustration of the excitation mechanism of the standing spin waves is shown in the inset of Fig.\ \ref{Fig:THzPrecessions}(a). As discussed earlier [inset Fig.\ \ref{Fig:KittelPrecessions}(a)], a short and intense transverse (OOP) spin current is injected into the top IP magnetized layer after the fs laser-pulse excitation. The spin current is absorbed very locally near the injection interface \cite{Lalieu2017}, creating a strong gradient in the OOP magnetization component in the top layer, as illustrated in the figure ($t = 0$). This highly non-equilibrium magnetization state relaxes by the excitation of (damped) higher-order standing spin waves, as illustrated for $n=0,1,2$ and $3$. In the following, only the first-order ($n=1$) standing spin wave is investigated. It is noted, however, that up to the third-order standing spin waves have been observed using a $20$ nm thick CoB absorption layer.

A typical measurement of the first-order standing spin wave is presented in Fig.\ \ref{Fig:THzPrecessions}(a), in which the measurement at a Co thickness of $13$ nm is shown. The observed dynamics is a superposition of two damped oscillations, which can be separated using a fit including two damped sines and a double exponential background (red solid line). The two fitted precessions are illustrated by the black ($0.15$ THz) and blue ($0.08$ THz) solid lines in the figure. Although the presence of two precessions could be explained by two different standing spin wave orders, it turns out that the slower precessions (blue curve) corresponds to an acoustic strain wave traveling along the depth of the multilayer. The acoustic strain wave is present in the polar MOKE measurement due to a lattice-deformation-induced change in the magneto-optical signal from the Co/Ni multilayer when the acoustic wave passes through it. A more detailed analysis of the acoustic strain wave can be found in Supplementary Note 2. 

The faster precession, indicated by the black curve, belongs to the (first-order) ferromagnetic standing spin wave. Measuring this precession at different positions along the Co wedge allows to extract the standing spin wave frequency $f_{\mathrm{sw}}$ as a function of the Co thickness, of which the result is presented in Fig.\ \ref{Fig:THzPrecessions}(b) (black dots). In this figure, also the result of the same measurement on the CoB sample is presented (blue dots). The dispersion relation for the standing spin waves is given by \cite{Lalieu2017} (using $M_{\mathrm{eff}}$)
\begin{multline}
f_{\mathrm{sw}}=\frac{\gamma}{2\pi}\left[\left(B_{\mathrm{app}}+\frac{D_{\mathrm{sw}}}{\gamma\hbar}k^{2}\right)\right. \\
\left.\times\left(B_{\mathrm{app}}+\mu_{0}M_{\mathrm{eff}}+\frac{D_{\mathrm{sw}}}{\gamma\hbar}k^{2}\right)\right]^{1/2},
\label{Eq:THzDispersion}
\end{multline}
in which $D_{\mathrm{sw}}$ corresponds to the spin wave stiffness, and the wave number $k$ of the $n^{\mathrm{th}}$ order standing spin wave is given by
\begin{equation}
k=\frac{\pi n}{t}.
\label{Eq:WaveNumber}
\end{equation}

The red solid lines in Fig.\ \ref{Fig:THzPrecessions}(b) are fits to the data using Eq.\ (\ref{Eq:THzDispersion}). The fits are done using the earlier obtained $g$ factor and thickness dependent $M_{\mathrm{eff}}$. Furthermore, with $B_{\mathrm{app}} = 0$ and $n=1$, this leaves $D_{\mathrm{sw}}$ as the only fitting parameter. As can be seen, the measured standing spin wave frequencies are well described by the dispersion relation. In case of the Co sample, however, a deviation from the dispersion curve can be seen around a Co thickness of $t_{\mathrm{Co}} \approx 7-10$ nm. The exact reason for this is not known. It is noted, however, that a change in crystallographic structure has been reported from the fcc structure for $t_{\mathrm{Co}} < 6$ nm to the hcp structure for bulk \cite{Tokac2015}. In case of the CoB top layer, which is known to be amorphous, such a crystallographic change should not be present. Looking at the measured dispersion for CoB (blue dots), it can be seen that there is no such deviation from the fitted dispersion curve. This suggests that the deviation seen for the Co top layer could indeed be related to the change in crystallographic structure. A more elaborate investigation should be performed (e.g., using XRD) to confirm this hypothesis.

The fitted values of the spin wave stiffness for the Co and CoB samples are $D_{\mathrm{sw}} = 882 \pm 8$ meV$\mathrm{\AA}^{2}$ and $D_{\mathrm{sw}} = 582 \pm 7$ meV$\mathrm{\AA}^{2}$, respectively. In the case of Co, experimental values for thin films ($<140$ nm) range between $250-520$ meV$\mathrm{\AA}^{2}$ \cite{Vaz2008}. Surprisingly, the measured value for the present Co sample is much higher. In the case of the (amorphous) CoB sample, the measured $D_{\mathrm{sw}}$ is also high compared to a (bulk) literature value of $\approx 170$ meV$\mathrm{\AA}^{2}$ \cite{Konc1994}. This suggests that the enhanced value of $D_{\mathrm{sw}}$ is not related to the crystalline structure. Moreover, the ratio of the measured values for Co and CoB is comparable to the ratio of the literature values. This indicates that the origin of the enhanced spin wave stiffness is the same for both used absorption layers. 

The large value of $D_{\mathrm{sw}}$ might be related to strain or intermixing at the absorption layer boundaries. In the case of Co, this can lead to lattice deformations. Since the spin wave stiffness is highly dependent on the lattice constant $a$ of the material ($D_{\mathrm{sw}} = 2 J S a^{2}$ with $J$ the exchange constant and $S$ the atomic spin), such lattice deformations are expected to have a significant influence on the spin wave stiffness. In case of the amorphous CoB, this effect would be present in the pair distribution function. It is also noted, without going into details, that the amplitude of the standing spin waves in the measured signal depends on the depth profile of the polar MOKE sensitivity within the absorption layer, which is known to be influenced by (amongst others) the attenuation of the laser and interface effects. If, for instance, the MOKE would be more sensitive to both interface regions and less to the bulk of the absorption layer, the (net) signal of the odd-order spin waves would be suppressed [see inset Fig.\ \ref{Fig:THzPrecessions}(a)]. In that case, the measured spin waves in Fig.\ \ref{Fig:THzPrecessions}(b) are the second-order standing spin waves ($n=2$), and the resulting spin wave stiffness for the Co layer would be $D_{\mathrm{sw}} \approx (882/4 =) 220$ meV$\mathrm{\AA}^{2}$. Although this value seems to be more in line with the literature values, the validity of such a MOKE-sensitivity-profile related suppression of the odd-order spin waves should be tested (especially for the thicker absorption layer thicknesses). Clearly, more research is needed in order to fully comprehend the enhanced value of the spin wave stiffness, for which the presented noncollinear bilayers could be of great value.

Next to the precession frequency, the damped sine fits of the standing spin waves [Fig.\ \ref{Fig:THzPrecessions}(a)] also provide the characteristic damping time $\tau_{\mathrm{sw}}$, which can be used to determine the Gilbert damping parameter $\alpha_{\mathrm{sw}}$ of the THz spin waves. The damping is calculated using 
\begin{equation}
\alpha_{\mathrm{sw}} = \left[\gamma\tau_{\mathrm{sw}} \left(B_{\mathrm{app}} + \frac{\mu_{0} M_{\mathrm{eff}}}{2}+ \frac{D_{\mathrm{sw}}}{\gamma \hbar} k^{2}\right)\right]^{-1},
\label{Eq:DampingTHz}
\end{equation}
which is similar as the equation used for the homogeneous precession [Eq.\ (\ref{Eq:DampKittel})], with an additional term resulting from the exchange interaction. 

\begin{figure}
	\includegraphics[scale=0.35]{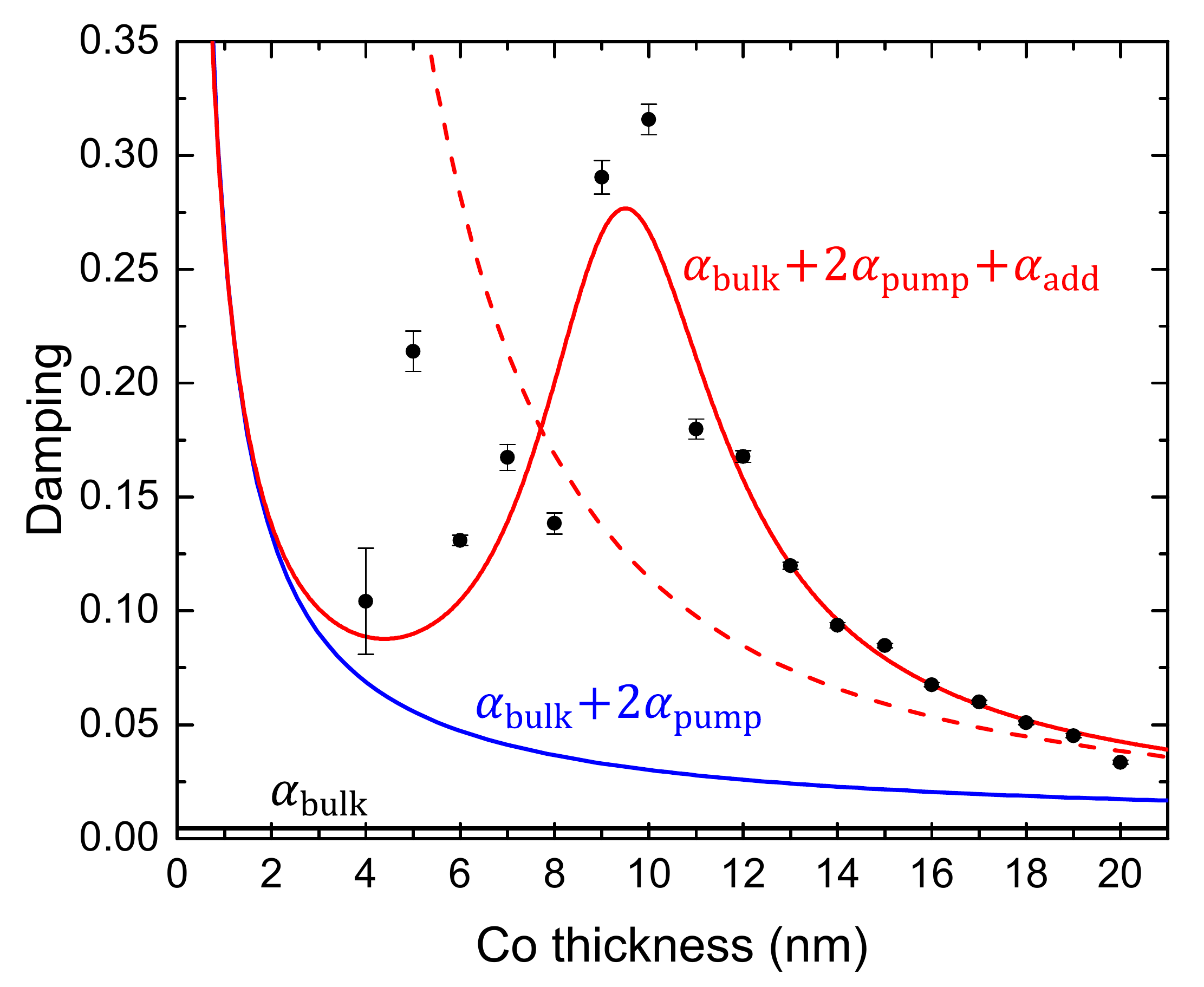}
	\caption{Gilbert damping for the higher-order standing spin waves as a function of the Co thickness. The $\alpha_{\mathrm{bulk}}$ and $2 \alpha_{\mathrm{pump}}$ contributions to the total damping are illustrated by the black and blue solid curves. The red solid line represents the fit to the data using Eq.\ (\ref{Eq:DampingSW}).}
	\label{Fig:THzDamping}
\end{figure}

The measured damping as a function of the Co layer thickness is presented in Fig.\ \ref{Fig:THzDamping} (black dots). Similar as for the homogeneous precessions, both the intrinsic damping ($\alpha_{\mathrm{bulk}}$) and interface spin pumping ($\alpha_{\mathrm{pump}}$) are contributing to the damping. In case of the (inhomogeneous) standing spin waves, the damping due to interface spin pumping is twice as large as for the homogeneous precession \cite{Kapelrud2013}. The $\alpha_{\mathrm{bulk}}$ and $2 \alpha_{\mathrm{pump}}$ contributions to the total damping are illustrated by the black and blue solid curves in the figure. Note that the values of $\alpha_{\mathrm{bulk}}$ and $\alpha_{\mathrm{pump}}$ are the ones determined from the homogeneous precessions [inset Fig.\ \ref{Fig:KittelPrecessions}(b)].

The figure clearly shows that there is an additional contribution to the damping $\alpha_{\mathrm{add}}$, which has a surprising thickness dependence, and enhances the damping up to about an order of magnitude compared to the damping of the homogeneous precession. The thickness dependence of $\alpha_{\mathrm{add}}$ can be divided into two regions. For $t_{\mathrm{Co}} \geq 10$ nm, a strong increase in $\alpha_{\mathrm{add}}$ is seen when decreasing the Co thickness. For $t_{\mathrm{Co}} < 10$ nm, the additional damping vanishes upon further reduction of the Co thickness. The same behavior was found in the CoB sample, which is shown in Supplementary Note 3. 

The additional source of damping might be the result of the inhomogeneity of the standing spin waves, for which an additional contribution to the damping was modelled in Ref.\ \cite{Tserkovnyak2009}. In this model, the additional damping originates from spin pumping between regions in the magnetic material that are precessing at a different phase. This damping term was calculated to scale with $k^{2}$, which is proportional to $t_{\mathrm{Co}}^{-2}$ in the present case. The damping based on this model (including $\alpha_{\mathrm{bulk}}$ and $2 \alpha_{\mathrm{pump}}$) is illustrated in the figure by the red dashed line. At first sight, the behavior of this additional source of damping does not seem to agree with the measurement. The $t_{\mathrm{Co}}^{-2}$ dependence does not include the reduction in damping for $t_{\mathrm{Co}}<10$ nm, and a thickness dependence much stronger than $t_{\mathrm{Co}}^{-2}$ is observed for $t_{\mathrm{Co}} \geq 10$ nm. 

The derivation of the additional damping in Ref.\ \cite{Tserkovnyak2009} was done for low frequencies, i.e., for slow dynamics, thereby neglecting frequency dependent terms in the transport equation for the spin current. Adding these terms in the derivation results in the following equation for the additional damping (see Supplementary Note 4 for the derivation)
\begin{equation}
\alpha_{\mathrm{add}} = A \ \mathrm{Re}\left[\frac{\tau_{\perp}\left(1+i\tau_{\perp} 2\pi f_{\mathrm{sw}}\right)}{\left(\tau_{\perp}\Delta_{\mathrm{xc}}/\hbar\right)^{2}-\left(-i+\tau_{\perp}2\pi f_{\mathrm{sw}}\right)^{2}}\right]k^{2}.
\label{Eq:DampingSW}
\end{equation}
In this equation, $A$ is a constant prefactor (discussed later), $\tau_{\perp}$ is the transverse spin scattering time, $\Delta_{\mathrm{xc}}$ the exchange energy and $f_{\mathrm{sw}}$ the precession frequency, given by the fit in Fig.\ \ref{Fig:THzPrecessions}(b). A fit to the data using this equation is shown by the red solid line in Fig.\ \ref{Fig:THzDamping}. From a qualitative point of view, it can be seen that the data is well described by Eq.\ (\ref{Eq:DampingSW}), which is also the case for the CoB sample (Supplementary Note 3). As can be seen in the figure, the extended model correctly describes the strong thickness dependence for $t_{\mathrm{Co}} \geq 10$ nm, where the increased dependence on $t_{\mathrm{Co}}$ with respect to the initial model (dashed red curve) results from the thickness dependence of $f_{\mathrm{sw}}$. Moreover, the extended model reproduces the reduction of the damping for $t_{\mathrm{Co}} < 10$ nm, reducing $\alpha_{\mathrm{add}}$ down to zero when $t_{\mathrm{Co}} \rightarrow 0$, i.e., for $f_{\mathrm{sw}} \rightarrow \infty$. In this high frequency limit, where $f_{\mathrm{sw}} \gg \tau_{\perp}^{-1}$, the angular momentum dissipation ($\propto \tau_{\perp}^{-1}$) becomes too slow, and its damping effect on the spin wave precession becomes negligible. 

A more quantitative analysis of the fit can be done by looking at the fitted values of $A$, $\tau_{\perp}$ and $\Delta_{\mathrm{xc}}$. For $A$, a value of $(3.1 \pm 0.2) \cdot 10^{-6}$ m$^{2}$ s$^{-1}$ is obtained. This prefactor is equal to \cite{Tserkovnyak2009}
\begin{equation}
A= \frac{n_{\mathrm{e}}\hbar^{2}}{4 m^{*} S},
\end{equation}
with $n_{\mathrm{e}}$ the electron number density, $\hbar$ the reduced Planck constant, $m^{*}$ the effective electron mass, and $S$ the spin density. Using the Drude conductivity $\sigma_{\mathrm{D}}$, which is given by $\sigma_{\mathrm{D}} = (n_{\mathrm{e}}e^{2}\tau_{\mathrm{D}})/m^{*}$, the spin wave density can be calculated using
\begin{equation}
S = \frac{\hbar^{2}\sigma_{\mathrm{D}}}{ 4 \tau_{\mathrm{D}} e^{2} A}.
\end{equation}
In this equation $e$ and $\tau_{\mathrm{D}}$ are the charge and mean free time of the electron, respectively. The spin density can in turn be used to calculate the amount of spins per Co atom. With a mean free path of $\approx 10$ nm and a Fermi velocity of $2.55 \cdot 10^{5}$ m s$^{-1}$ in Co\cite{Gall2016}, the mean free time is equal to $\tau_{\mathrm{D}} \approx 39$ fs. Together with a conductivity of $\sigma_{\mathrm{D}} = 1.79 \cdot 10^{7}$ S m$^{-1}$ in Co \cite{Kasap2017}, and the assumption of an fcc lattice with a lattice constant of $a_{0} = 3.54 \mathrm{\AA}$ \cite{Liu1996}, a spin density of $1.69 \, \hbar$ per Co atom is calculated. This value is close to the known value of $1.72$ for Co, and thereby supports the validity of the fit.

The transverse spin scattering time was found to be $\tau_{\perp} = 1.5 \pm 0.2$ ps. This scattering time is related to the disorder scattering time $\tau_{\mathrm{dis}}$ and electron-electron scattering time $\tau_{\mathrm{ee}}$ via \cite{Tserkovnyak2009}, $\tau_{\perp}^{-1} = \tau_{\mathrm{dis}}^{-1} + \tau_{\mathrm{ee}}^{-1}$. Unfortunately, no corresponding values for Co were found in the literature. Lastly, the fitted exchange energy is equal to $\Delta_{\mathrm{xc}} = 0.93 \pm 0.04$ meV. This is much lower than the exchange energy known for the $d$ electrons in Co, which is in the order of $0.1$ eV. However, the fitted exchange interaction might need to be compared to the exchange energy for the $s$ electrons at the Fermi surface, which is expected to be much smaller. 

\section{Conclusion}
In conclusion, it has been demonstrated that the noncollinear magnetic bilayer is a convenient tool to excite and investigate THz standing spin waves, thereby showing high potential for future THz magnonics. Using wedge-shaped absorption layers, the spin wave dispersion in Co and CoB was measured. Analysis of the dispersion resulted in a surprisingly high spin wave stiffness for both materials, for which further investigation is needed in order to clarify the enhanced values. Additionally, the noncollinear magnetic bilayers were used to investigate the damping of the THz standing spin waves, demonstrating a large damping contribution, additional to the bulk damping and damping resulting from interface spin pumping. The additional damping displayed a strong increase of the damping with decreasing absorption layer thickness, and a previously unseen sudden decrease in the damping for the thinnest films. A model for the additional damping contribution was proposed. The observed decrease in the (additional) damping for the highest spin wave frequencies might be of great relevance for future magnonics, in which high frequency spin waves with low damping are desired. 

\acknowledgments{We thank Y. Tserkovnyak for valuable discussions. This work is part of the Gravitation program 'Research Centre for Integrated Nanophotonics', which is financed by the Netherlands Organisation for Scientific Research (NWO).}

\subsection*{\large Supplementary Note 1: Analysis homogeneous precession for the CoB sample}

For the analysis of the THz standing spin waves, the effective magnetization and Gilbert damping parameter (bulk and interface spin pumping contributions) are needed as a function of the thickness of the absorption layer. These properties are determined using the homogeneous precession, of which the analysis for the noncollinear magnetic bilayer with a Co absorption layer is shown in the main paper. In this section, the results of similar measurements performed on the sample with the CoB absorption layer are presented. 

\begin{figure}
	\includegraphics[scale=0.35]{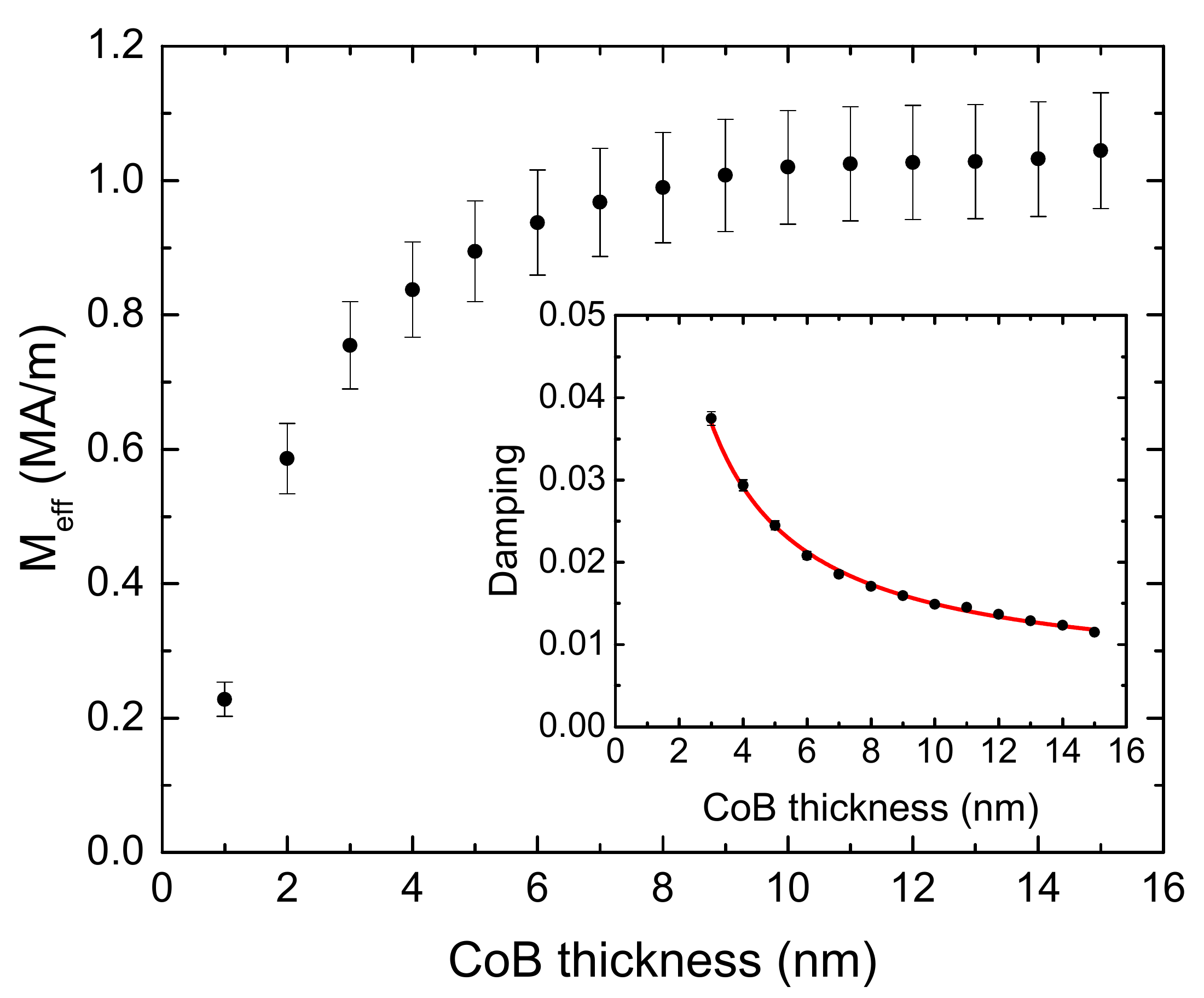}
	\caption{Effective magnetization $M_{\mathrm{eff}}$ as a function of the CoB thickness. The inset shows the Gilbert damping parameter as a function of the CoB thickness, in which the damping determined with the different magnetic field amplitudes are averaged. The red curve represents a fit to the data using Eq.\ (3) of the main paper.}
	\label{Fig:CoBKittelPrecessions}
\end{figure}

As was mentioned in the main paper, the effective magnetization $M_{\mathrm{eff}}$ of the absorption layer at a certain thickness is determined by measuring the frequency $f_{\mathrm{IP}}$ of the homogeneous precession as a function of the applied magnetic field $B_{\mathrm{app}}$, and using the Kittel relation [Eq.\ (1) of the main paper] to fit the value of $M_{\mathrm{eff}}$ (and $\gamma$). Figure\ \ref{Fig:CoBKittelPrecessions} presents the resulting effective magnetization as a function of the CoB thickness $t_{\mathrm{CoB}}$. The observed thickness dependence of $M_{\mathrm{eff}}$ results from an out-of-plane surface anisotropy, which decreases $M_{\mathrm{eff}}$, and of which the contribution falls off as $t_{\mathrm{CoB}}^{-1}$. The obtained thickness dependent $M_{\mathrm{eff}}$ is later used in the analysis of the THz standing spin waves (Fig.\ 2(b) of the main paper, and Supplementary Note 3). The Kittel fits also allows the determination of the $g$ factor using the fitted value of $\gamma$. For the CoB sample, a $g$ factor of $2.31 \pm 0.08$ was found, which is similar as the one found for the Co absorption layer. 

Using the characteristic damping time $\tau$ of the homogeneous precessions, together with the previously determined value of $M_{\mathrm{eff}}$, the Gilbert damping constant $\alpha$ at each CoB thickness can be determined, using Eq.\ (2) of the main paper. The measured damping as a function of the CoB thickness is shown in the inset of Fig.\ \ref{Fig:CoBKittelPrecessions}, in which the damping determined with the different magnetic field amplitudes are averaged. The fitted values for the bulk damping and interface spin pumping amplitude are $\alpha_{\mathrm{bulk}} = (5.5 \pm 0.2) \cdot 10^{-3}$ and $A_{\mathrm{pump}} = (0.94 \pm 0.02) \cdot 10^{-10}$ m, respectively. Both values are used later when evaluating the damping of the THz standing spin waves (Supplementary Note 3).

\subsection*{\large Supplementary Note 2: Acoustic strain waves}

The precession measurement of the first-order standing spin wave in the Co sample (at a thickness of $t_{\mathrm{Co}} = 13$ nm) presented in Fig.\ 2(a) of the main paper showed a superposition of two precessions. The second slower precession of $0.08$ THz was attributed to a longitudinal acoustic strain wave traveling along the depth of the multilayer. In this section a more elaborate analysis of the slow precessions is presented, which demonstrates that it indeed belongs to a laser-induced acoustic strain wave.

\begin{figure}
	\includegraphics[scale=0.35]{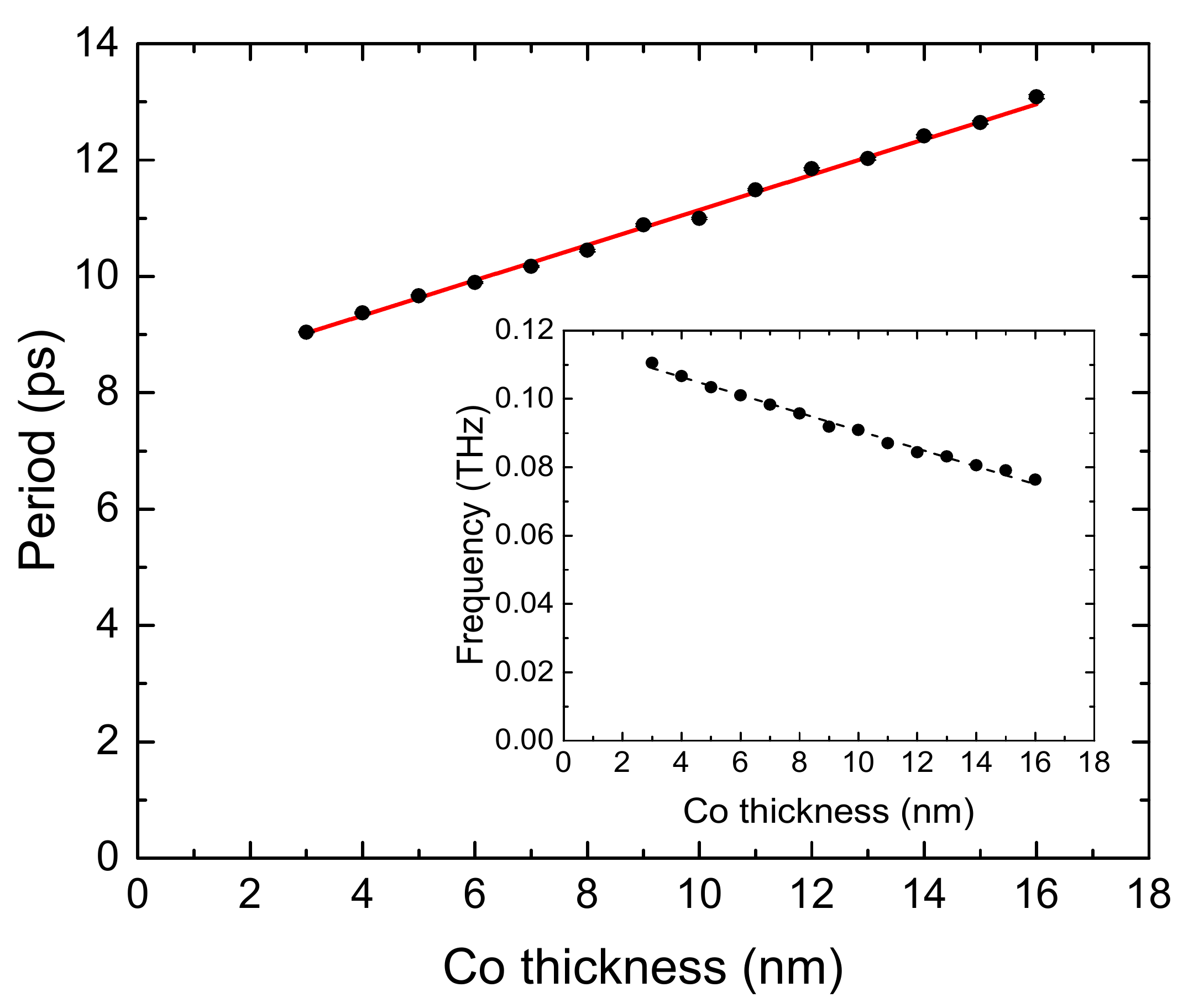}
	\caption{Precession period as a function of the Co thickness for the second ($0.08$ THz) precession measured in Fig.\ 2(a) of the main paper (blue curve). The solid red curve is a linear fit to the data. The inset shows the precession frequency as a function of the Co thickness, where the black dotted line is a guide to the eye.}
	\label{Fig:AcousticFrequency}
\end{figure}

The frequency of the precession as a function of Co thickness is displayed in the inset of Fig.\ \ref{Fig:AcousticFrequency}. At first sight, due to the Co thickness dependency, it appears that the precession exists in the Co layer. However, (in a different measurement) it turned out that the precession was observed along the Co wedge down to a thickness of $t_{\mathrm{Co}} = 0$, i.e., without the Co layer. Moreover, the decrease in frequency with increasing Co thickness appears to be close to linear (dotted line, guide to the eye), which neither fits the Co thickness dependence of the fundamental precession nor the higher-order standing spin wave dispersion. The observed behavior does fit with a (laser-induced) longitudinal acoustic strain wave that travels through the full multilayer. For such a wave, the period $p$ of one round trip is given by
\begin{equation}
	p = \frac{2 t}{v_{\mathrm{l}}},
	\label{Eq:PeriodAC}
\end{equation}
in which $t$ is the thickness of the multilayer, and $v_{\mathrm{l}}$ is the longitudinal sound velocity. 

To check that the measured precession indeed belongs to the acoustic strain wave, the frequency data in the inset of Fig.\ \ref{Fig:AcousticFrequency} is converted to the precession period as a function of the Co thickness, which is presented in the main figure and fitted using a linear fit. Looking at Eq.\ \ref{Eq:PeriodAC}, it can be seen that the fitted slope is equal to $2/v_{\mathrm{l,Co}}$, with $v_{\mathrm{l,Co}}$ the longitudinal sound velocity in Co. The resulting sound velocity is $v_{\mathrm{l,Co}} \approx 6.6$ km s$^{-1}$. This value is close to the literature value of $5.7$ km $^{-1}$ (Ref.\ \cite{Martienssen2005}). Moreover, measurements on a similar noncollinear magnetic bilayer with a wedge-shaped Co absorption layer resulted in a sound velocity of $v_{\mathrm{l,Co}} \approx 5.5$ km s$^{-1}$. Lastly, using a wedge-shaped Pt layer instead of the Co absorption layer, a sound velocity in Pt of $v_{\mathrm{l,Pt}} \approx 3.6$ km s$^{-1}$ was found, again close to the literature value of $4.08$ km s$^{-1}$ (Ref.\ \cite{Martienssen2005}). 

In conclusion, the analysis presented in this section show that the second ($0.08$ THz) precession observed in Fig.\ 2(a) of the main paper correspond to the a laser-induced longitudinal acoustic strain wave. Although such waves have been measured before using time-resolved measurements of the reflectivity, here, they are measured in the magneto-optical signal. It is believed that this is due to a lattice-deformation-induced change in the magneto-optical signal coming from the Co/Ni multilayer (bottom OOP magnetic layer in the noncollinear magnetic bilayer) when the acoustic strain wave passes through it. One of the reasons for this conclusion is that the sign of the precession was seen to invert when the magnetization in the Co/Ni multilayer was reversed. 

\subsection*{\large Supplementary Note 3: Damping THz standing spin waves in the CoB absorption layer}

In the main paper, the damping of the THz standing spin waves in the Co absorption layer was investigated (Fig.\ 3 of the main paper), which demonstrated an additional contribution to the damping $\alpha_{\mathrm{add}}$ (on top of the bulk damping $\alpha_{\mathrm{bulk}}$ and the damping resulting from interface spin pumping $2\alpha_{\mathrm{pump}}$). Moreover, the additional damping displayed an unexpected thickness dependence, showing a strong increase in $\alpha_{\mathrm{add}}$ with decreasing Co thickness for $t_{\mathrm{Co}} \geq 10$ nm, while the additional damping vanished upon further reduction of the Co thickness for $t_{\mathrm{Co}} < 10$ nm. In this section, it is demonstrated that the same additional damping is present in case of the CoB absorption layer, with the same thickness dependence.

\begin{figure}
	\includegraphics[scale=0.35]{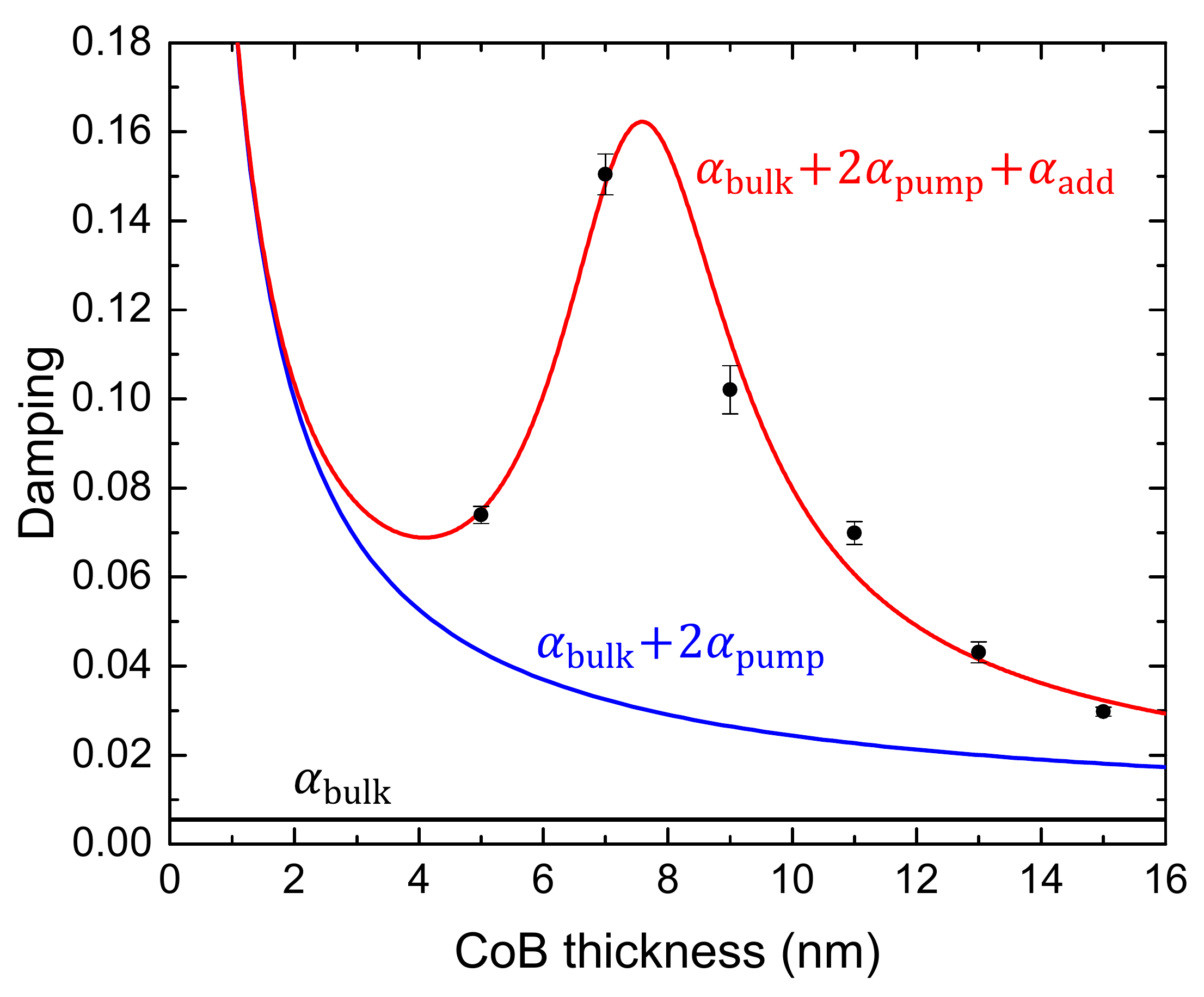}
	\caption{Gilbert damping for the higher-order standing spin waves as a function of the CoB thickness. The $\alpha_{\mathrm{bulk}}$ and $2 \alpha_{\mathrm{pump}}$ contributions to the total damping are illustrated by the black and blue solid curves. The red solid line represents the fit to the data using Eq.\ (7) of the main paper.}
	\label{Fig:CoBTHzDamping}
\end{figure}

The measured damping as a function of the CoB layer thickness is presented in Fig.\ \ref{Fig:CoBTHzDamping} (black dots). The $\alpha_{\mathrm{bulk}}$ and $2 \alpha_{\mathrm{pump}}$ contributions to the total damping are illustrated by the black and blue solid curves in the figure. Note that the values of $\alpha_{\mathrm{bulk}}$ and $\alpha_{\mathrm{pump}}$ are the ones determined from the homogeneous precessions in Supplementary Note 1. The red solid line represents the fit to the data using Eq.\ (7) of the main paper. As can be seen, the same dependence of the additional damping on the magnetic layer thickness is found as for the Co sample in Fig.\ 3 of the main paper, including a strong increase in damping for $t_{\mathrm{CoB}} \geq 8$ nm, and a decrease in damping for $t_{\mathrm{CoB}} < 8$ nm. 

\subsection*{\large Supplementary Note 4: Derivation THz additional damping equation}
In this section, the equation for the additional damping $\alpha_{\mathrm{add}}$ as given in Eq.\ (7) of the main paper is derived, which is an extension to the more elaborate derivation given in Ref.\ \cite{Tserkovnyak2009}. For simplicity, only the exchange interaction is taken into account in the effective field, i.e., there is no applied field and no anisotropy contributions, and the canting of the magnetization away from its equilibrium direction ($+y$) is assumed to be small. As a result, the magnetization can be described by a precession in the $x,z$ plane. With the inclusion of a $k \, z$ phase term to allow standing spin waves in the $z$ direction, the normalized magnetization is described by
\begin{equation}
	\vec{m} = \begin{bmatrix}
		m_{x} \\[0.3em]
		m_{y} \\[0.3em]
		m_{z}
	\end{bmatrix}
	= \begin{bmatrix}
		m_{0} \mathrm{e}^{i \left(k z + \omega t\right)} 	\\[0.3em]
		1                          						\\[0.3em]
		i m_{0} \mathrm{e}^{i \left(k z + \omega t\right)}
	\end{bmatrix},
	\label{Eq:Magn}
\end{equation}
where the real parts represent the physical components of the magnetization. The effective field $\vec{H}_{\mathrm{ex}}$ is equal to 
\begin{equation}
	\vec{H}_{\mathrm{eff}} = \frac{2 A_{\mathrm{ex}}}{\mu_{0} M_{\mathrm{s}}} \nabla^{2} \vec{m},
	\label{Eq:EffField}
\end{equation}
in which $M_{\mathrm{s}}$ and $A_{\mathrm{ex}}$ are the saturation magnetization and exchange stiffness of the material, respectively. 

Following Ref.\ \cite{Tserkovnyak2009}, and omitting the Gilbert damping term, the LLG equation including the additional spin-wave damping term is given by
\begin{equation}
	\frac{d\vec{m}}{dt} = -\gamma \mu_{0} \left(\vec{m} \times \vec{H}_{\mathrm{eff}} \right) - \frac{\eta(\omega)}{S}\left(\vec{m} \times \nabla^{2}\frac{d\vec{m}}{dt} \right),
	\label{Eq:LLGSW}
\end{equation}
in which $S$ is the spin density and $\eta(\omega)$ a phenomenological parameter characterizing spin-wave damping, which can be dependent on the precession frequency $\omega$. Evaluating either the $x$ or $z$ component, the equation can be solved for the precession frequency
\begin{equation}
	\omega = \frac{2 \gamma A_{\mathrm{ex}}}{M_{\mathrm{s}}}k^{2} + \frac{\eta(\omega)}{S}k^{2} i \omega.
	\label{Eq:OmegaSW}
\end{equation}
The first term on the right-hand side can be recognized as the standard (exchange) spin-wave frequency given by $\omega_{\mathrm{sw}} = D_{\mathrm{sw}}/\hbar \, k^{2}$, in which the relation between $A_{\mathrm{ex}}$ and the spin wave stiffness $D_{\mathrm{sw}}$, as given by\cite{Coey2010}
\begin{equation}
	A_{\mathrm{ex}} = \frac{M_{\mathrm{s}} D_{\mathrm{sw}}}{2 \gamma \hbar},
\end{equation}
is used. Looking at the second term, it can be seen that the imaginary part of $\eta(\omega)$ alters the precessions frequency, while the real part contributes to the damping. Using a first order approximation, in which the contribution of $\eta(\omega)$ to the precession frequency ($\mathrm{Im}\left[\eta(\omega_{\mathrm{sw}})\right]$) is considered small, and therefore can be neglected, the precession frequency can be rewritten to 
\begin{equation}
	\omega \simeq \omega_{\mathrm{sw}} + \frac{\mathrm{Re}\left[\eta(\omega_{\mathrm{sw}})\right]}{S}k^{2} \, i \omega_{\mathrm{sw}}.
	\label{Eq:OmegaSWReal}
\end{equation}
The imaginary term on the right-hand side corresponds to the spin-wave damping, in which the additional spin-wave damping parameter is given by
\begin{equation}
	\alpha_{\mathrm{add}} = \frac{\mathrm{Re}\left[\eta(\omega_{\mathrm{sw}})\right]}{S}k^{2}.
	\label{Eq:DampingSWReal}
\end{equation}

From Ref.\ \cite{Tserkovnyak2009} [Eqs.\ (22) and (28)], the (frequency independent) phenomenological spin-wave damping parameter is given by
\begin{equation}
	\eta = \frac{n_{\mathrm{e}}\hbar^{2}}{4 m^{*}}\frac{\tau_{\perp}}{1+\left(\tau_{\perp}\Delta_{\mathrm{xc}}/\hbar\right)^{2}},
	\label{Eq:DampSWPaper}
\end{equation}
with $n_{\mathrm{e}}$ the electron number density, $\hbar$ the reduced Planck constant, $m^{*}$ the effective electron mass, $\tau_{\perp}$ the transverse spin scattering time and $\Delta_{\mathrm{xc}}$ the exchange energy. This derivation was performed for low frequency, thereby neglecting the frequency dependent term in the transport equation for the spin current [Eq.\ (24)]. The frequency dependence can be included by using the substitution
\begin{equation}
	\frac{1}{\tau_{\perp}} \rightarrow \frac{1}{\tau_{\perp}} + i \omega_{\mathrm{sw}}.
\end{equation}
Finally, combining this substitution with Eqs.\ (\ref{Eq:DampingSWReal}) and (\ref{Eq:DampSWPaper}), and using $\omega_{\mathrm{sw}} = 2 \pi f_{\mathrm{sw}}$, the fit equation for the additional damping as shown in Eq.\ (7) of the main paper is obtained, 
\begin{equation}
	\alpha_{\mathrm{add}} = A \ \mathrm{Re}\left[\frac{\tau_{\perp}\left(1+i\tau_{\perp} 2\pi f_{\mathrm{sw}}\right)}{\left(\tau_{\perp}\Delta_{\mathrm{xc}}/\hbar\right)^{2}-\left(-i+\tau_{\perp}2\pi f_{\mathrm{sw}}\right)^{2}}\right]k^{2},
	\label{Eq:DampSW}
\end{equation}
in which
\begin{equation}
	A= \frac{n_{\mathrm{e}}\hbar^{2}}{4 m^{*} S}.
\end{equation}

\end{document}